\documentclass[a4paper]{article}
\usepackage{Odyssey2020}
\usepackage{epsfig,amssymb,amsmath}
\ninept

\usepackage{multirow}

\usepackage{cite}

\title{Bayesian x-vector: Bayesian Neural Network based x-vector System for Speaker Verification}

\name{Xu Li$^{\star}$, Jinghua Zhong$^{\star,\dagger}$, Jianwei Yu$^{\star}$, Shoukang Hu$^{\star}$, Xixin Wu$^{\star}$, Xunying Liu$^{\star}$, Helen Meng$^{\star}$}

\address{$^{\star}$Department of Systems Engineering and Engineering Management, \\
                   The Chinese University of Hong Kong \\
         $^{\dagger}$SpeechX Limited, Shenzhen, China\\
{\small \tt \{xuli, jwyu, skhu, wuxx, xyliu, hmmeng\}@se.cuhk.edu.hk, jhzhong@speechx.cn}}
\begin{document}
\maketitle

\begin{abstract}
Speaker verification systems usually suffer from the mismatch problem between training and evaluation data, such as speaker population mismatch, the channel and environment variations. In order to address this issue, it requires the system to have good generalization ability on unseen data.
In this work, we incorporate Bayesian neural networks (BNNs) into the deep neural network (DNN) x-vector speaker verification system to improve the system's generalization ability. With the weight uncertainty modeling provided by BNNs, we expect the system could generalize better on the evaluation data and make verification decisions more accurately.
Our experiment results indicate that the DNN x-vector system could benefit from BNNs especially when the mismatch problem is severe for evaluations using out-of-domain data. Specifically, results show that the system could benefit from BNNs by a relative EER decrease of 2.66\% and 2.32\% respectively for short- and long-utterance in-domain evaluations. Additionally, the fusion of DNN x-vector and Bayesian x-vector systems could achieve further improvement. Moreover, experiments conducted by out-of-domain evaluations, e.g. models trained on Voxceleb1 while evaluated on NIST SRE10 core test, suggest that BNNs could bring a larger relative EER decrease of around 4.69\%.

\end{abstract}


\section{Introduction}
\label{sec: intro}

We are observing an ever-increasing use of automatic speaker verification (ASV) systems in our everyday lives.
An essential step for verification is to disentangle the speaker information from each spoken utterance and then decisions are made based on the speaker similarity. Through decades of years development, three most representative frameworks have been proposed in this research area. (i) Extending from joint factor analysis \cite{kenny2007joint,kenny2007speaker} that models speaker and channel subspaces separately, i-vector based speaker embedding is proposed to jointly model speaker and channel variations together with a speaker-discriminative back-end for decisions. Such systems include Gaussian mixture model (GMM) i-vector \cite{dehak2010front,kenny2012small,prince2007probabilistic,garcia2011analysis} and deep neural network (DNN) i-vector systems \cite{lei2014novel}. (ii) Benefiting from the powerful discrimination ability of DNNs, DNN-based speaker embedding is proposed to extract speaker-discriminative representations for each utterance, which could perform as the state of the art on short-utterance evaluation conditions. These systems include d-vectors \cite{variani2014deep} and x-vectors \cite{snyder2017deep,snyder2018x}. (iii) With the development of end-to-end techniques, many researches focus on constructing ASV systems in an end-to-end manner \cite{zhang2016end,heigold2016end,snyder2016deep}, which directly learns a mapping from enrollment and testing utterance pairs to verification scores, resulting in a compact structure and comparably good performance.

A challenging issue for ASV systems development is the mismatch between the training and evaluation data, such as the speaker population mismatch, and variations in channel and environmental background. The speaker population used for training and evaluation commonly have no overlap especially for practical applications. To overcome this mismatch usually requires the extracted speaker representations to generalize well on unseen speaker data. 
The channel and environment variations mostly exist in practical applications where the training and evaluation data are collected from different types of recorders and environments. These mismatches also have a high demand for the model's generalizability on unseen data.

To address this issue, previous efforts \cite{wang2018unsupervised,bhattacharya2019adapting,tu2019variational} have applied adversarial training to alleviate the channel and environment variations from utterance embedding. It is achieved by adding adversarial penalty on domain-related information in embedding vectors during the extractor training stage. 
This approach has been proven to be effective in alleviating the effects of channel and environmental mismatches. However, it does not consider the speaker population mismatch that could also lead to the system performance degradation. 
In this work, we try to improve the system's generalizability across these kinds of mismatches in a unified approach. Inspired by previous work based on Bayesian learning \cite{xie2019blhuc,hu2019bayesian,yu2019comparative}, we focus on the DNN x-vector system and apply Bayesian neural networks (BNNs) to improve the x-vector system's generalization ability.

The Bayesian learning approach has been shown to be effective to improve the generalization ability of discriminative training in DNN systems.
In the machine learning community, similar work have been conducted to incorporate Bayesian learning into DNN systems.
Barber et al. \cite{barber1998ensemble} proposed an efficient variational inference strategy for BNNs. Blundell et al. \cite{blundell2015weight} proposed a novel backpropagation-compatible algorithm for learning the network parameters' posterior distribution. 
In the speech area, some previous work involved BNNs in speech recognition \cite{graves2011practical,hu2019bayesian,xie2019blhuc,hu2019lf}. Especially, Xie et al. \cite{xie2019blhuc} proposed the Bayesian learning of hidden unit contributions (BLHUC) for speaker adaptation. The BLHUC could model the uncertainty of speaker-dependent parameters and improve the speech recognition performance especially when given very limited speaker adapatation data. Other work also applied the Bayesian technique into language modelling \cite{chien2015bayesian,yu2019comparative}.

In a DNN x-vector system, the parameters of traditional time delay neural network (TDNN) layers estimated via the maximum likelihood strategy are deterministic and tend to overfit when given limited training data or when there is a large mismatch between the training and evaluation data. In the case of mismatch in speaker population, the overfitted model parameters may result in speaker representations following a spike distribution towards possible training speaker identities. However this will tend not to generalize well on unseen speaker data. To address this issue, BNNs could help smooth the distributions of speaker representation for better generalization on unseen speaker data.

The cases of channel and environmental mismatch are similar. For instance, for channel mismatch, the overfitted model parameters may partially rely on channel information to classify speakers due to various recorders for different speakers in the training data. 
However, when generalizing to channel-mismatched evaluation data, the original channel-speaker relationship is broken and the trained reliance on channel information cloud lead to misclassification.
To alleviate this issue, BNNs change deterministic parameters to be probabilistic via a posterior distribution. This parameter distribution modeling could reduce the risk of overfitting on channel information by smoothing parameters to consider extra possible values that do not rely on channel information for speaker classification.



The above issues motivate this work to incorporate BNNs into the x-vector system by replacing the TDNN layers. We adopt an efficient variational inference based approach to approximate the parameter posterior distribution. The effectiveness of Bayesian learning is investigated on both short- and long-utterance in-domain evaluation, and also an out-of-domain evaluation that includes larger channel and environment mismatches. To the best of our knowledge, this is the first work that applies Bayesian learning technique to speaker verification systems.

Our experiments are based on Voxceleb1 (for short-utterance condition) and NIST Speaker Recognition Evaluation (SRE) 10 (for long-utterance condition) datasets. 

The rest of this paper is organized as follows: Section~\ref{sec: baseline} introduces the baseline system architecture, and the Bayesian neural networks will be illustrated in Section~\ref{sec: BNN}. Section~\ref{sec: exp-setup} states the experimental setups, and Section~\ref{sec: exp-results} shows the experimental results. Finally, Section~\ref{sec: conclusion} concludes the paper.

\section{Baseline: DNN x-vector system}
\label{sec: baseline}

A DNN x-vector system \cite{snyder2018x} consists of two parts: a front-end used for extracting utterance-level speaker embeddings and a verification scoring back-end. The front-end compresses speech utterances of different length into fixed-dimension speaker-related embeddings (x-vectors). Based on these embeddings, different scoring schemes can be used for judging whether two utterances belong to a same person or not. In this work, we focus on the reversion of the front-end, and choose different back-ends for the performance evaluation.

\begin{figure}[t]
\includegraphics[width=\columnwidth]{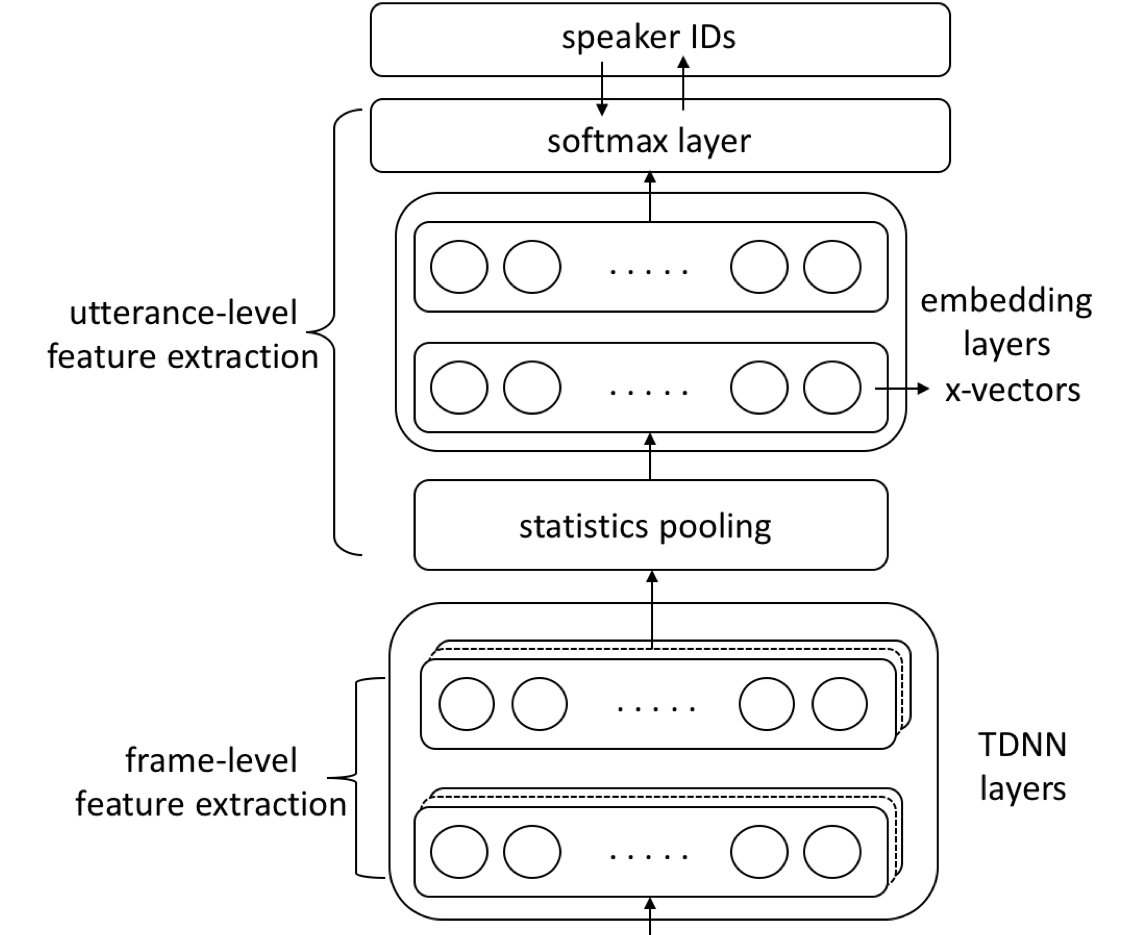}
\caption{The architecture of a DNN x-vector extractor.}
\label{fig: x-vector-system}
\end{figure}

The x-vector extractor is a neural network trained via a speaker discrimination task, the architecture of which is shown in Fig.~\ref{fig: x-vector-system}. It consists of frame-level and utterance-level extractors. At the frame level, several layers of time delay neural network (TDNN) are used to model the time-invariant characteristics of acoustic features. Then the statistics pooling layer aggregates all the frame-level outputs from the last TDNN layer, and computes their mean and standard deviation. The mean and standard deviation are concatenated together and propagated through several fully connected utterance-level layers, i.e. embedding layers, and finally the softmax output layer. The cross-entropy between one-hot speaker labels and the softmax outputs is used as the loss function during the training stage. In the testing stage, given the acoustic features of an utterance, the embedding layer output is extracted as the x-vector. Since the network is trained in a speaker-discriminative manner, the extracted x-vectors are expected to only contain speaker-related information. But in practice, as investigated in \cite{raj2019probing}, x-vectors still contain other speaker-unrelated information, such as channel, transcription and utterance-length. These information could affect the verification performance especially on the mismatched evaluation data.

\section{Bayesian neural network}
\label{sec: BNN}

\subsection{Weight uncertainty}
Traditional neural networks learn a set of deterministic parameters to fit with the training data via the maximum likelihood estimation, and then make inference based on these fixed parameters in the testing stage. This estimation may lead to overconfident parameters when the training data is limited or when there exists a mismatch between the training and evaluation data. 

To alleviate this issue, Bayesian neural networks learn the parameters' posterior distribution instead. The posterior distribution $p(w|\mathcal{D})$ based on the training data $\mathcal{D}$ models weight uncertainty and theoretically enables an infinite number of possible model parameters to fit with the data. This weight uncertainty modeling can smooth model parameters and make the model generalize well on unseen data. During the testing stage, the model computes the output $\hat{y}$ given the input {$x$} by making an expectation over the weight posterior distribution $P(w|\mathcal{D})$, as shown in Eq.~\ref{eq:compute-output}. 


\begin{align}
    \nonumber
    p(\hat{y}|x) &= E_{p(w|\mathcal{D})}[p(\hat{y}|x,w)] \\
                 &= \int p(\hat{y}|x,w) p(w|\mathcal{D}) dw \label{eq:compute-output} 
\end{align}

%

The estimation of the weight posterior distribution $P(w|\mathcal{D})$ is an essential procedure when training a Bayesian neural network. However, the direct estimation is intractable for neural networks of any practical size, since the number of possible weight values could be infinity. So the variational approximation \cite{barber1998ensemble} is commonly adopted to estimate the weight posterior distribution.

\subsection{Variational approximation for Bayesian learning}
The variational approximation estimates a set of parameters $\theta_q$ for a distribution $q(w;\theta_q)$ to approximate the posterior distribution $p(w|\mathcal{D})$. This is achieved by minimizing the Kullback-Leibler (KL) divergence between these two distributions, as shown in Eq.~\ref{eq: variational-approximation-initial}. 

\begin{align}
    \nonumber
    \theta_q^{\star} &= \arg\min_{\theta_q} KL(q(w;\theta_q)||p(w|\mathcal{D})) \\
    &= \arg\min_{\theta_q} \int q(w;\theta_q)\log \frac{q(w;\theta_q)}{p(w|\mathcal{D})} dw \label{eq: variational-approximation-initial} \\
    &= \arg\min_{\theta_q} \int q(w;\theta_q)\log \frac{q(w;\theta_q)}{p(w)p(\mathcal{D}|w)} dw + \log p(\mathcal{D}) \label{eq: variational approximation-0} \\
    &= \arg\min_{\theta_q} \int q(w;\theta_q)\log \frac{q(w;\theta_q)}{p(w)p(\mathcal{D}|w)} dw \label{eq: variational approximation-1} \\
    \nonumber
    &= \arg\min_{\theta_q} \int q(w;\theta_q) \log \frac{q(w;\theta_q)}{p(w)} dw \\
    &~~~~~~~~~~~~~~~~~- \int q(w;\theta_q) \log p(\mathcal{D}|w) dw \label{eq: variational approximation-2} \\
    &= \arg\min_{\theta_q} KL(q(w;\theta_q)||p(w)) - E_{q(w;\theta_q)}[\log(p(\mathcal{D}|w))]
    \label{eq: varitional approximation-3}
\end{align}

From Eq.~\ref{eq: variational-approximation-initial} to \ref{eq: variational approximation-1}, we apply Bayes' Rule and drop the constant term $\log p(\mathcal{D})$ that does not affect the minimization over $\theta_q$. Equations from \ref{eq: variational approximation-1} to \ref{eq: varitional approximation-3} demonstrate that this minimization equation could be decomposed into two parts: 1) the KL divergence between the approximation distribution $q(w;\theta_q)$ and the prior distribution $p(w)$ on the weight, 2) the expectation of the log likelihood of the training data over the approximation distribution $q(w;\theta_q)$. Eq.~\ref{eq: varitional approximation-3} is used as the loss function to be minimized during the training process.

As commonly adopted in \cite{barber1998ensemble,blundell2015weight}, we assume that the variational approximation follows a diagonal Gaussian distribution with a parameter set $\theta_q=\{\mu_q, \rho_q\}$. $\mu_q$ is the mean of the diagonal Gaussian distribution, while $\rho_q$ generates the diagonal Gaussian standard deviation $\sigma_q$ by $\sigma_q=\log(1+\exp(\rho_q))$. The prior distribution is also assumed to be a diagonal Gaussian distribution with a parameter set $\theta_p=\{\mu_p, \sigma_p\}$. Unlike $\theta_q$ will be updated during the training stage, $\theta_p$ is usually a set of predetermined fixed parameters. 

Under the Gaussian distribution assumptions, the first part in Eq.~\ref{eq: varitional approximation-3} has a closed-form result that can be computed directly,

\begin{align}
    \label{eq: KL-part}
    \nonumber
    &KL(q(w;\theta_q)||p(w)) \\
    &= \sum_{d=1}^D [ \log(\frac{\sigma_{p,d}}{\sigma_{q,d}})+\frac{(\mu_{q,d}-\mu_{p,d})^2+\sigma_{q,d}^2}{2\sigma_{p,d}^2}-\frac{1}{2}]
\end{align}
where $D$ denotes the number of entries in the weight matrix, and $\mu_{q,d}$, $\sigma_{q,d}$, $\mu_{p,d}$ and $\sigma_{p,d}$ are the $d$-th entry of $\mu_{q}$, $\sigma_{q}$, $\mu_{p}$ and $\sigma_{p}$, respectively.

While the integration in the second part cannot be computed directly, Monte Carlo sampling is commonly applied to approximate this integration, as shown in Eq.~\ref{eq: monte-carlo-sampling}:

\begin{align}
    \label{eq: monte-carlo-sampling}
    \nonumber
    E_{q(w;\theta_q)}[\log(p(\mathcal{D}|w))] &= \int q(w;\theta_q) \log(p(\mathcal{D}|w)) dw \\
                                  &\approx \frac{1}{J} \sum_{j=1}^J \log(p(\mathcal{D}|w^j))
\end{align}
where $J$ is the number of samples and $w^j=\mu_q+\sigma_q\bigodot\epsilon_j$ is the $j$ th sample from the distribution $q(w;\theta_q)$, and $\bigodot$ denotes the element-wise multiplication. As the equation shows, $w^j$ is sampled by scaling and shifting a random signal $\epsilon_j \thicksim N(0,I)$ from the unit Gaussian distribution.

Finally, the loss function is derived as:
\begin{align}
    \label{eq: loss-function}
    \nonumber
    &L = KL(q(w;\theta_q)||p(w)) - E_{q(w;\theta_q)}[\log(p(\mathcal{D}|w))] \\
    \nonumber
    &\approx \sum_{d=1}^D [ \log(\frac{\sigma_{p,j}}{\sigma_{q,j}})+\frac{(\mu_{q,j}-\mu_{p,j})^2+\sigma_{q,j}^2}{2\sigma_{p,j}^2}-\frac{1}{2}] \\
    &~~~~- \frac{1}{J} \sum_{j=1}^J \log(p(\mathcal{D}|w^j))
\end{align}

The gradient with respective to parameters $\theta_q=\{\mu_q, \rho_q\}$ can be derived as,

\begin{align}
    \label{eq: gradients}
    &\frac{\partial L}{\partial \mu_{q,d}} = \frac{\mu_{q,d}-\mu_{p,d}}{\sigma_{p,d}^2}+\frac{1}{J}\sum_{j=1}^J G_{j,d} \\
    \frac{\partial L}{\partial \rho_{q,d}} = [&\frac{\sigma_{q,d}}{\sigma_{p,d}^2}-\frac{1}{\sigma_{q,d}}] \frac{e^{\rho_{q,d}}}{1+e^{\rho_{q,d}}} + \frac{1}{J} \sum_{j=1}^J \epsilon_{j,d} G_{j,d} \frac{e^{\rho_{q,d}}}{1+e^{\rho_{q,d}}}
\end{align}
where $G_{j,d}=-\frac{\partial \log(p(\mathcal{D}|w^j))}{\partial w^j_d}$ is the standard gradient of loss function with respective to $w^j_d$ (the $d$-th entry of the weight matrix).

\begin{table}[t]
\centering
\begin{tabular}{c|c|c}
\hline
 Layer & Layer Context & Input $\times$ Output \\
 \hline
 frame1 & [$t$-2, $t$+2] & 120 $\times$ 512 \\
 \hline
 frame2 & \{$t$-2, $t$, $t$+2\} & 1536 $\times$ 512 \\
 \hline
 frame3 & \{$t$-3, $t$, $t$+3\} & 1536 $\times$ 512 \\
 \hline
 frame4 & \{$t$\}           & 512 $\times$ 512 \\
 \hline
 frame5 & \{$t$\}           & 512 $\times$ 1500 \\
 \hline
 stats pooling & [0, $T$) & 1500$T$ $\times$ 3000 \\
 \hline
 segment6 & \{0\}        & 3000 $\times$ 512 \\
 \hline
 segment7 & \{0\}        & 512 $\times$ 512 \\
 \hline
 softmax & \{0\} & 512 $\times$ $N$ \\
 \hline
\end{tabular}
\caption{The x-vector extractor architecture. The $N$ in the softmax layer is the size of speaker population during training. The $t$ in frame-level layers represents the current frame, and $T$ represents the total number of frames in an utterance. x-vectors are extracted from segment6, before the nonlinearity.}
\label{tab: x-vector-sys}
\end{table}

\begin{table*}[]
\caption{In-domain evaluation: equal error rate (EER) and minimum detection cost function (min-DCF) in different conditions.}
\label{tab:indomain-eer-dcf}
\centering
\begin{tabular}{c|c|c|c|c|c|c}
\hline
\hline
 Training set & Evaluation set & System & Scoring back-end & x-vector extractor & EER(\%) & $DCF_{VOX}$/$DCF_{SRE10}$ \\
\hline
\multirow{6}{*}{Voxceleb1} & \multirow{6}{*}{Voxceleb1} & (1) & \multirow{3}{*}{cosine} & baseline & 9.58          & 0.6899 \\
                          &                            & (2) &                         & proposed & 9.30          & 0.6508 \\
                          &                            & (3) &                         & fusion   & \textbf{8.64} & \textbf{0.6423} \\
\cline{3-7}
                          &                            & (4) & \multirow{3}{*}{PLDA}   & baseline & 6.68          & 0.6023 \\
                          &                            & (5) &                         & proposed & 6.52          & \textbf{0.5423} \\
                          &                            & (6) &                         & fusion   & \textbf{6.35} & 0.5487 \\
\hline
\hline
\multirow{6}{*}{NIST SRE10} & \multirow{6}{*}{NIST SRE10} & (7) & \multirow{3}{*}{cosine} & baseline & 5.61          & 0.6830 \\
                            &                             & (8) &                         & proposed & 5.52          & 0.6555 \\
                            &                             & (9) &                         & fusion   & \textbf{5.47} & \textbf{0.6502} \\
\cline{3-7}
                            &                             & (10) & \multirow{3}{*}{PLDA}   & baseline & 3.29          & 0.3926 \\
                            &                             & (11) &                         & proposed & 3.19          & \textbf{0.3835} \\
                            &                             & (12) &                         & fusion   & \textbf{3.17} & 0.3840 \\
\hline
\hline
\end{tabular}
\end{table*}

\section{Experimental setup}
\label{sec: exp-setup}

In order to evaluate the effectiveness of Bayesian learning for speaker verification in both short- and long-utterance conditions, we perform experiments on two datasets. For the short-utterance condition, we consider the Voxceleb1 \cite{nagrani2017voxceleb} dataset, where the recordings are short clips of human speech. There are totally 148,642 utterances for 1,251 celebrities. We follow the configuration in \cite{nagrani2017voxceleb}, where 4,874 utterances from 40 speakers are reserved for evaluation, and the remaining utterances are used for training x-vector systems (the baseline and BNN-based system) and the back-end model parameters. 

For the long-utterance condition, the core test in the NIST speaker recognition evaluation 10 (SRE10) \cite{martin2010nist} is used for evaluation, where the recordings are long-duration telephone speech. The training data used in this condition includes SREs from 04 through 08, Switchboard2 Phases 1, 2 and 3, and Switchboard cellular. The Switchboard portion is commonly used to increase the training data variety \cite{snyder2017deep,snyder2018x}. In total there are around 65,000 recordings for 6,500 speakers. All this training data is used for training x-vector systems (the baseline and BNN-based system), while only the SRE parts are used for training the scoring back-ends. During the training stage, since the utterances are very long, the GPU memories limitation forces a tradeoff between minibatch size and maximum training example length. We randomly cut each recording into chunks of length from 2s to 10s (200 frames to 1000 frames) along with a minibatch size of 48. After this procedure, the average training chunks for each speaker is around 220. No data augmentation technique is applied in any experiment.

Mel-frequency cepstral coefficients (MFCCs) are adopted as acoustic features in all the experiments. Before extracting MFCCs, a pre-emphasis with coefficient of 0.97 is adopted. ``Hamming" window having size of 25ms and step-size of 10ms is applied to extract a frame, and finally 30 cepstral coefficients are kept. 
The extracted MFCCs are mean-normalized over a sliding window of up to 3 seconds, and voice activity detection (VAD) \cite{snyder2017deep} filters out nonspeech frames.


The configuration of the baseline x-vector extractor is consistent with \cite{snyder2018x}, as shown in Table~\ref{tab: x-vector-sys}. After propagating through several frame-level layers and one statistic pooling layer, the outputs from the first segment-level layer (segment6 in the table) before nonlinearity are extracted as x-vectors. We adopt the stochastic gradient descent (SGD) optimizer during the training stage. In order to make a fair comparison, the Bayesian x-vector system is configured with the same architecture of the baseline system except the first TDNN layer is replaced by BNN layer with the same number of units. We also attempted to replace other layers with BNN layer, but experiment results show that operation on other layers gives a slightly worse performance than operation on the first layer. This may indicate that operation on the layer close to the input features is more effective and has more impact on improving the model's generalization ability. The choice of prior distribution has an impact on model convergence and training efficiency. In this work, we set the prior distribution based on the baseline model parameters, similar with the strategy in \cite{hu2019bayesian}. The x-vector systems (the baseline and BNN-based system) are implemented by Pytorch \cite{NEURIPS2019_9015}, while the other parts, including data preparation, feature extraction and training scoring back-ends, are implemented by Kaldi toolkit \cite{Povey_ASRU2011}.

To evaluate the generalization benefits that Bayesian learning could bring under evaluation of different mismatch degrees, we design two kinds of evaluation experiments: in-domain evaluation and out-of-domain evaluation. The training and testing stages are executed on the same dataset in in-domain evaluation, while they are executed on different datasets in out-of-domain evaluation. On both evaluations, we perform experiments on two datasets, i.e. Voxceleb1 for the short-utterance condition and NIST SRE10 for the long-utterance condition. Two kinds of scoring back-ends are adopted in our experiments: cosine and probabilistic linear discriminative analysis (PLDA) back-ends. Before propagating into the back-end scoring, speaker embeddings are projected into a lower dimension space via linear discriminant analysis (LDA). Following the default settings adopted in Kaldi toolkit \cite{Povey_ASRU2011}, the LDA dimension is set as 150 and 200 for cosine and PLDA scoring, respectively.

The evaluation metrics adopted in this work are the commonly used equal error rate (EER) and minimum detection cost function (min-DCF). For the min-DCF metric, we consider the prior probability of target trials to be 0.01 on the Voxceleb1 (denoted as $DCF_{VOX}$). For the NIST SRE10 dataset, the target trial partitions are much smaller and thus we consider the prior probability to be 0.001 (denoted as $DCF_{SRE10}$). Intuitively, with the consideration that the baseline system could have correct operation with high confidence on the common characterisitcs between the training and evaluation data, while the BNN-based system could generate well on the mismatch characteristics, we also design a fusion system for performance comparison. The fusion is operated by averaging the verification scores from the two systems.

\begin{table*}[]
\caption{Out-of-domain evaluation: equal error rate (EER) and minimum detection cost function (min-DCF) in different conditions.}
\label{tab:out-of-domain-eer-dcf}
\centering
\begin{tabular}{c|c|c|c|c|c|c}
\hline
\hline
 Training set & Evaluation set & System & Scoring back-end & x-vector extractor & EER(\%) & $DCF_{VOX}$/$DCF_{SRE10}$ \\
\hline
\multirow{6}{*}{Voxceleb1} & \multirow{6}{*}{NIST SRE10} & (1) & \multirow{3}{*}{cosine} & baseline & 10.78         & 0.8650 \\
                           &                             & (2) &                         & proposed & 10.38          & 0.8633 \\
                           &                             & (3) &                         & fusion   & \textbf{10.15} & \textbf{0.8428} \\
\cline{3-7}
                            &                             & (4) & \multirow{3}{*}{PLDA}   & baseline & 8.31          & 0.8646 \\
                            &                             & (5) &                         & proposed & 7.84          & 0.8541 \\
                            &                             & (6) &                         & fusion   & \textbf{7.71} & \textbf{0.8378} \\
\hline
\hline
\multirow{6}{*}{NIST SRE10} & \multirow{6}{*}{Voxceleb1} & (7) & \multirow{3}{*}{cosine} & baseline & 15.30          & 0.8101 \\
                            &                            & (8) &                         & proposed & 14.85          & 0.8164 \\
                            &                            & (9) &                         & fusion   & \textbf{14.14} & \textbf{0.7913} \\
\cline{3-7}
                            &                            & (10) & \multirow{3}{*}{PLDA}   & baseline & 11.27          & 0.7636 \\
                            &                            & (11) &                         & proposed & 10.91          & 0.7555 \\
                            &                            & (12) &                         & fusion   & \textbf{10.68} & \textbf{0.7461} \\
\hline
\hline
\end{tabular}
\end{table*}

\section{Experiment results}
\label{sec: exp-results}

\begin{figure}
    \centering
    \includegraphics[width=\columnwidth]{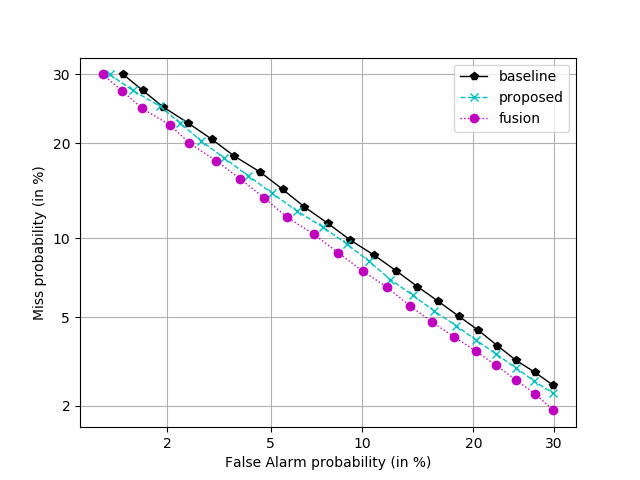}
    \caption{In-domain evaluation: detection error trade-off (DET) curves on the Voxceleb1 dataset, using a cosine scoring backend.}
    \label{fig:DET-indomain}
\end{figure}

\subsection{In-domain evaluation}
In this section, we perform in-domain evaluation on two datasets: Voxceleb1 for the short-utterance condtion and NIST SRE10 for the long-utterance condition.
The corresponding performance is shown in Table~\ref{tab:indomain-eer-dcf}. From the table, we observe that EERs consistently decrease after incorporating the Bayesian learning in both short- and long-utterance conditions. In each condition, we consider the average relative EER decrease across cosine and PLDA back-ends.
In the short-utterance condition, the average relative EER decrease from Bayesian x-vector system is 2.66\%, and the fusion system could achieve further average relative EER decrease by 7.24\%. For the long-utterance condition, the average relative EER decrease is 2.32\% for Bayesian x-vector system and 3.08\% for the fusion system.

Our experiment results show that systems in the short-utterance condition could benefit more from the Bayesian learning when compared with the long-utterance condition. One possible explanation is that, in the short-utterance condition, systems may be heavily affected by speaker-unrelated information, such as channel and phonetic information, which may bring larger mismatches in the testing stage. With the uncertainty modeling of model parameters, the Bayesian learning could bring extra benefits to alleviate these larger mismatches and improve the performance.
Similar results could be observed in detection cost function (DCF) metrics as shown in the last column of Table~\ref{tab:indomain-eer-dcf}. Fig.~\ref{fig:DET-indomain} illustrates the detection error trade-off (DET) curves of systems with the cosine back-end (Systems 1, 2 and 3 in Table~\ref{tab:indomain-eer-dcf}). It shows that the proposed Bayesian system outperforms the baseline for all operating points, and the fusion system could achieve further improvements due to the complementary advantages of the baseline and the Bayesian system.


\begin{figure}
    \centering
    \includegraphics[width=\columnwidth]{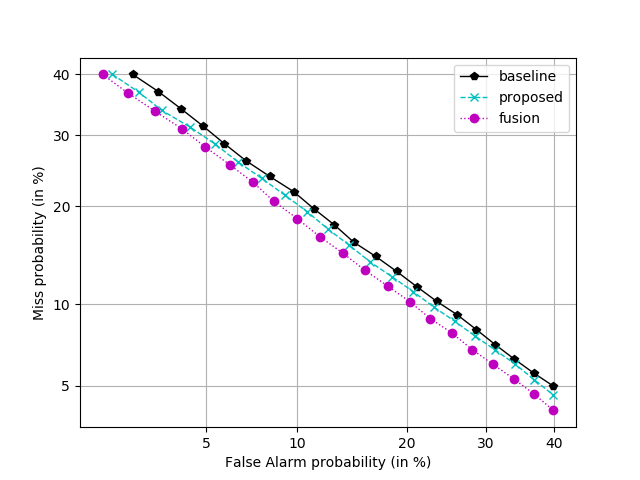}
    \caption{Out-of-domain evaluation: detection error trade-off (DET) curves on the Voxceleb1 dataset, using a cosine scoring backend.}
    \label{fig:DET-out-of-domain}
\end{figure}

\subsection{Out-of-domain evaluation}
The out-of-domain evaluation is performed in this section, as shown in Table~\ref{tab:out-of-domain-eer-dcf}. The model trained on Voxceleb1 (Systems 1 to 6) will be evaluated on NIST SRE10, and vice versa. System performance usually drops significantly due to the larger mismatch between the training and evaluation data, so this evaluation has a higher demand for the system's generalization ability.

From Table~\ref{tab:out-of-domain-eer-dcf}, we observe that systems could benefit more from the generalization power of Bayesian learning. We also consider the average relative EER decrease across cosine and PLDA scoring back-ends for performance evaluation. In the experiments evaluated on NIST SRE10, the average relative EER decrease is 4.69\% and 6.53\% for the Bayesian system and the fusion system, respectively. For the experiments on the Voxceleb1 dataset, the average relative EER decrease is 3.07\% for the Bayesian x-vector system, and the fusion system achieves a further average relative EER decrease of 6.41\%.
The larger relative EER decrease compared with that in in-domain evaluation suggests that Bayesian learning could be more beneficial when larger mismatch exists between the training and evaluation data. This phenomenon is similar to the observations stated in \cite{hu2019lf}, where the improvement on the out-of-domain dataset (with larger mismatch) is larger than the in-domain dataset. The last column in Table~\ref{tab:out-of-domain-eer-dcf} shows the corresponding DCF performance, and we observe consistent improvement by applying Bayesian learning and the fusion system. Similar to the observations in Fig.~\ref{fig:DET-indomain}, the DET curves in Fig.~\ref{fig:DET-out-of-domain} show consistent improvements by applying Bayesian learning and the fusion model for all operating points.



\section{Conclusion}
\label{sec: conclusion}
In this work, we incorporate the BNN technique into the DNN x-vector system to improve the model's generalization ability. BNN layers embedded in the x-vector extractor make the extracted speaker embedding (Bayesian x-vector) generalize better on unseen data. Our experimental results show that the DNN x-vector could benefit from Bayesian learning for both short- and long-utterance conditions, and the fusion system could achieve further performance improvements. Moreover, we observe that systems could benefit more from Bayesian learning in out-of-domain evaluation. Especially, in out-of-domain evaluation performed on the NIST SRE10 dataset, the average relative EER decrease across cosine and PLDA scoring is around 4.69\% and 6.53\% by applying the Bayesian system and the fusion system, respectively. This suggests that Bayesian learning is more beneficial when larger mismatch exists between the training and evaluation data. Possible future research will focus on incorporating Bayesian learning into the end-to-end speaker verification systems.

\section{Acknowledgements}
\label{sec: acknowledgements}
This work is partially supported by a grant from the HKSAR Government's Research Grants Council General Research Fund (reference number 14208718).

\bibliographystyle{IEEEbib}
\bibliography{main}

%

\end{document}